\documentclass[]{spie}  

\usepackage{amsmath,amsfonts,amssymb}
\usepackage{subcaption}
\usepackage{graphicx}
\usepackage{cite} 
\usepackage{times}
\usepackage{epsfig}
\usepackage{amsmath}
\usepackage{nccmath}
\usepackage{amssymb}
\usepackage{mwe}
\usepackage{acro}
\usepackage{amssymb}
\usepackage{xcolor,colortbl}
\usepackage{tabularx}
\usepackage{relsize}
\usepackage{pifont}
\usepackage{booktabs} 
\usepackage{multirow}
\usepackage{multicol}
\usepackage{adjustbox}
\usepackage{float}
\usepackage{graphicx}
\usepackage{makecell}
\usepackage{tabu}
\usepackage[colorlinks=true, allcolors=blue]{hyperref}
\usepackage[capitalize]{cleveref}

\title{Polyhedra Encoding Transformers: Enhancing Diffusion MRI Analysis Beyond Voxel and Volumetric Embedding}
   
\author[a]{Tianyuan Yao}
\author[b]{Zhiyuan Li}
\author[a]{Praitayini Kanakaraj}
\author[c]{Derek B. Archer}
\author[d]{Kurt Schilling}
\author[e]{Lori Beason-Held}
\author[e]{Susan Resnick}
\author[a,b,c,d]{Bennett A. Landman}
\author[a,d]{Yuankai Huo}

\affil[a]{Department of Computer Science, Vanderbilt University, Nashville, TN, USA}

\affil[b]{Department of Electrical and Computer Engineering, Vanderbilt University, Nashville, TN, USA}

\affil[c]{Department of Neurology, Vanderbilt University Medical Center, Nashville, TN, USA}

\affil[d]{Department of Biomedical Engineering, Vanderbilt University, Nashville, TN, USA}

\affil[e]{Laboratory of Behavioral Neuroscience, National Institute on Aging, Baltimore, MD, USA}

\begin{document} 
\maketitle

\begin{abstract}
Diffusion-weighted Magnetic Resonance Imaging (dMRI) is an essential tool in neuroimaging. It is arguably the sole noninvasive technique for examining the microstructural properties and structural connectivity of the brain. Recent years have seen the emergence of machine learning and data-driven approaches that enhance the speed, accuracy, and consistency of dMRI data analysis. However, traditional deep learning models often fell short, as they typically utilize pixel-level or volumetric patch-level embeddings similar to those used in structural MRI, and do not account for the unique distribution of various gradient encodings. In this paper, we propose a novel method called Polyhedra Encoding Transformer (PE-Transformer) for dMRI, designed specifically to handle spherical signals. Our approach involves projecting an icosahedral polygon onto a unit sphere to resample signals from predetermined directions. These resampled signals are then transformed into embeddings, which are processed by a transformer encoder that incorporates orientational information reflective of the icosahedral structure. Through experimental validation with various gradient encoding protocols, our method demonstrates superior accuracy in estimating multi-compartment models and Fiber Orientation Distributions (FOD), outperforming both conventional CNN architectures and standard transformers.

\end{abstract}


\section{INTRODUCTION}
\label{sec:intro}  
Diffusion MRI (dMRI) is one of the most important medical imaging tools and the only noninvasive approach that can probe tissue microstructures by tracking the restricted diffusion of water molecules in biological tissues~\cite{mori2006principles}. While the widely used diffusion tensor model has proven sensitive to pathological changes such as stroke and tumors~\cite{le2001diffusion}, it lacks specificity to microstructural properties like cell size, axonal diameter, fiber density, and orientational dispersion. To address this, advanced dMRI models have been developed to characterize specific microstructural features~\cite{novikov2019quantifying}, including intravoxel incoherent motion (IVIM)\cite{le1988separation}, neurite orientation dispersion and density imaging (NODDI)\cite{zhang2012noddi}, and soma and neurite density imaging (SANDI)\cite{palombo2020sandi}. These advanced models typically involve multiple compartments with complex, highly non-linear signal representations, making them prone to estimation errors when fitted using conventional non-linear optimization techniques such as non-linear least squares. Additionally, the acquisition of data required by these models, which includes multiple b-values and diffusion directions in q-space, is time-consuming and susceptible to motion artifacts—a significant limitation when imaging moving subjects such as abdominal organs, fetuses, and placentas.

To mitigate estimation errors and accelerate the acquisition process for advanced dMRI models, various methods have been proposed. For instance, Nedjati-Gilani et al.\cite{nedjati2017machine} introduced a random forest method to estimate microstructural parameters within the Kärger model\cite{karger1988principles}. The advent of deep learning techniques has further opened new avenues for dMRI model fitting. Traditional estimation methods, which typically fit diffusion signals on a voxel-wise basis, fail to consider the correlation between signals in neighboring voxels and overlook the spatial regularity of diffusion tensor values. To address this, the concept of q-space deep learning (q-DL)\cite{golkov2016q} was first proposed to directly map dMRI signals to diffusion kurtosis imaging (DKI) parameters using a subset of q-space data (with a reduced number of b-values and diffusion directions), employing a simple three-layer multilayer perceptron (MLP). Subsequent studies have explored different architectures: Gibbons et al.\cite{gibbons2019simultaneous} utilized a 2D convolutional neural network to estimate NODDI and generalized fractional anisotropy maps simultaneously, Koppers et al.\cite{koppers2019spherical} employed a residual network to enhance the comparability of dMRI signals acquired from different scanners. Additionally, Barbieri et al.\cite{barbieri2020deep} used a three-layer MLP with a self-supervised method to estimate IVIM model parameters. Beyond these end-to-end mapping approaches, model-driven neural networks, which introduce domain knowledge as prior information, have been proposed to enhance network performance and interpretability\cite{gregor2010learning, yang2018admm, wang2020model}. These networks are designed to unfold the optimization process of a mathematical model within the network architecture~\cite{liang2019deep}. Unlike conventional networks, model-driven networks are both data-driven and incorporate a model prior, making them more interpretable~\cite{wang2020model}. This approach has gained increasing popularity in medical imaging. For instance, Ye et al.\cite{ye2017tissue} introduced a model-based neural network for estimating NODDI parameters, and Zheng et al.\cite{zheng2021model} developed a model-driven sparsity coding deep neural network (SCDNN) for IVIM parameter estimation in the fetal brain.

However, convolution-based networks in current model-driven frameworks have a fixed receptive field within a single layer~\cite{luo2016understanding}, and stacking deeper convolutional layers leads to bloated models with sharply increasing computational loads~\cite{wang2018non}. To address this, a self-attention mechanism ~\cite{vaswani2017attention}that adapts to a dynamic receptive field can be integrated into the q-space deep learning task, forming a Transformer-like structure. Due to its superior performance and flexibility, the Transformer model has garnered significant interest across various fields. In image processing, the Vision Transformer (ViT)\cite{dosovitskiy2020image} has been introduced for classification tasks in computer vision, outperforming traditional convolutional networks. Despite its advantages, the application of ViT in medical imaging has been limited, primarily due to its high demand for training data. Unlike traditional models that incorporate inductive biases\cite{dosovitskiy2020image}, ViT requires large quantities of data for effective training. Previous deep-learning approaches have not sufficiently explored methods to leverage data from diverse gradient encoding schemes. As a result, many studies have focused on single datasets with uniform gradient encoding schemes or have relied on intermediate representations like spherical harmonics~\cite{nath2019enabling}. These approaches can limit model generalizability by failing to capture the full range of q-space signals, leading to suboptimal performance when applied to datasets with varying encoding schemes.

\begin{figure*}[t]
\begin{center}

\includegraphics[width=0.8\linewidth]{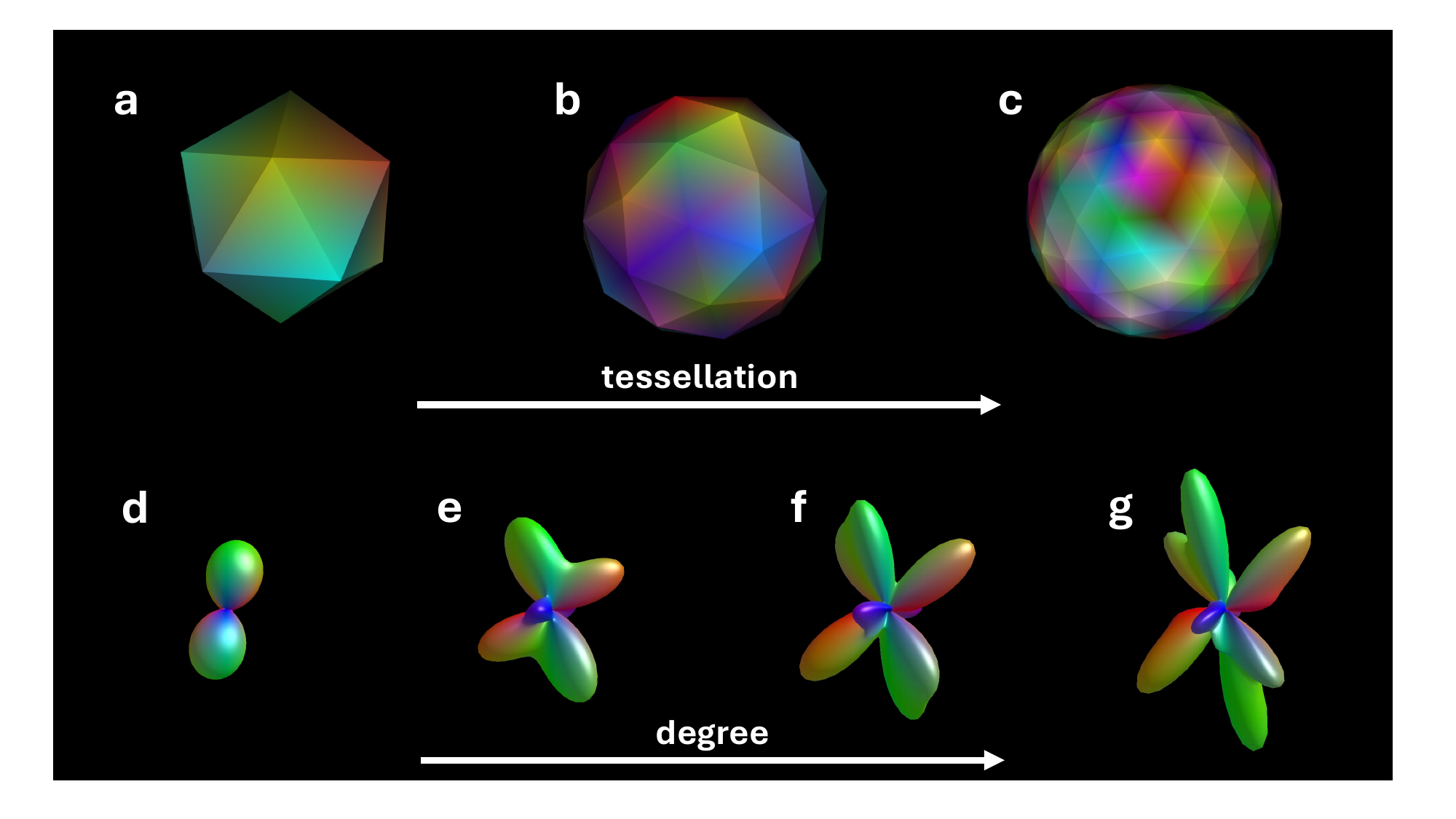}
\end{center}
\caption{For the Icosahedral polyhedron family (a,b,c), Icosahedral gradient encoding offers a compelling advantage for diffusion MRI by providing a nearly uniform and isotropic sampling of gradient directions on the unit sphere. The larger ICOSA sets (\(N_e > 6\)) can be generated by edge bisection and tessellation of the ICOSA6 directions and projection onto a unit sphere for diffusion MRI signal. For spherical harmonics (d,e,f,g shows the fiber ODF one same voxel with 2,4,6 and 8 degrees of spherical harmonics), Higher degrees (e.g., l=8) allow for more detailed representations, but there is a trade-off between capturing detail and introducing noise, as higher-order harmonics may fit noise in the data.}
\label{idea}
\end{figure*}

To bridge the gap between traditional approaches and the integration of Transformer models in diffusion MRI, it is crucial to explore methods that effectively utilize data from diverse gradient encoding schemes without relying on intermediate representations like spherical harmonics. While spherical harmonics offer a flexible and continuous representation, they can be computationally demanding and sensitive to noise, especially at higher degrees. Icosahedral polygon resampling~\cite{hasan2001comparison} presents a promising alternative, providing a uniform and isotropic distribution of sampling points on the sphere. This method reduces directional biases and offers robust signal representation, even in the presence of noise. To address the need for large amounts of data for training the transformer and to enhance the model interpretability. We applied transformer models to this structured and uniformly sampled q-space, we can harness the power of self-attention mechanisms to capture intricate patterns and dependencies across the diffusion signals. This approach not only maintains the rotational symmetry and isotropic properties inherent to the icosahedral sampling but also enables the model to learn spatial relationships in the diffusion data in a way that mirrors how ViT models learn from visual data. This could potentially enhance the model's ability to analyze complex gradient encoding patterns, leading to more accurate and robust diffusion MRI analysis. The contribution of this paper is threefold:

$\bullet$ The proposed PE-transformer leverages the strengths of Vision Transformers (ViT) by incorporating a novel configuration that uses icosahedral resampling to provide a uniform and isotropic sampling of q-space data, enhancing the model's ability to capture complex diffusion patterns.

$\bullet$ It is designed to be universally applicable for performing diffusion properties' estimation across heterogeneous single-shell diffusion MRI datasets, providing a solution that generalizes across different scanning protocols and datasets.

$\bullet$ 3D patches are employed to provide complete spatial information to enhance the performance.

\section{METHOD}
\label{sec:method}  

\begin{figure*}[t]
\begin{center}

\includegraphics[width=0.8\linewidth]{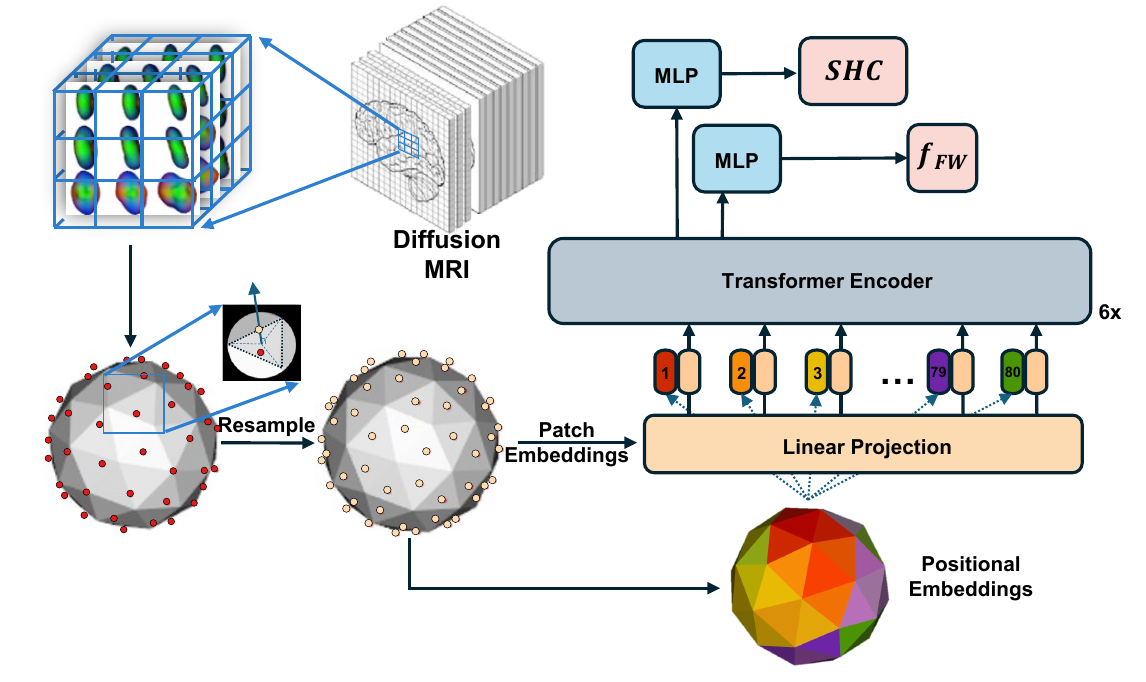}
\end{center}
\caption{The overall design of our proposed PE-Transformer.}
\label{model}
\end{figure*}

\subsection{Icosahedral (ICOSA) Polyhedra}
The regular icosahedron polyhedra family is defined by three polygon edges per face with five faces intersecting at each vertex~\cite{croft2012unsolved}. Face triangulation can provide sets with \(N_e = 5n^2 + 1\) for \(n = 1, 2, 3, \dots\)~\cite{tegmark1996icosahedron,sibley1998geometric }. This results in ICOSA6, ICOSA21, ICOSA46, ICOSA81, etc. The ICOSA6, ICOSA21 and ICOSA46 encoding schemes are illustrated in Fig. 1. The larger ICOSA sets (\(N_e > 6\)) can be generated by edge bisection and tessellation of the ICOSA6 directions and projection onto a unit sphere~\cite{croft2012unsolved}. Muthupalli et al.~\cite{muthupallai1999navigator} concluded that ICOSA6 is the optimum encoding scheme for DTI. The icosahedron has also been applied in encoding for chemical shielding tensor spectroscopy~\cite{conturo1995diffusion}. Tuch et al.~\cite{tuch1999high} used the ICOSA126 for diffusion measurements with high angular sampling.

\subsection{Polyhedra signal resampling}

To achieve uniform and isotropic resampling of diffusion MRI signals, we employed an icosahedral-based approach wherein the diffusion signal is projected onto the center points of an icosahedron's faces. Let \(\mathbf{g}_i \in \mathbb{R}^3\) denote the original gradient directions, and \(S(\mathbf{g}_i)\) the corresponding diffusion signals. We seek to resample these signals onto a new set of uniformly distributed directions \(\mathbf{r}_j \in \mathbb{R}^3\), which are the centroids of the faces of an icosahedron.

The resampled signal \(S'(\mathbf{r}_j)\) at each icosahedral direction \(\mathbf{r}_j\) is obtained using linear interpolation of the original signal values:

\begin{equation}
S'(\mathbf{r}_j) = \sum_{i=1}^{N} w_{ij} S(\mathbf{g}_i)
\end{equation}

\noindent where \(w_{ij}\) are the interpolation weights, calculated based on the proximity of \(\mathbf{r}_j\) to the original directions \(\mathbf{g}_i\). The weights \(w_{ij}\) are determined by the linear interpolation function:

\begin{equation}
w_{ij} = \frac{1}{\|\mathbf{r}_j - \mathbf{g}_i\|}
\end{equation}

\noindent subject to the normalization condition:

\begin{equation}
\sum_{i=1}^{N} w_{ij} = 1
\end{equation}



This approach ensures that the diffusion signal is uniformly resampled across the unit sphere, with the icosahedral points providing an isotropic and balanced set of directions for subsequent analyses.

\subsection{Polyhedra Encoding Transformer}
Following polyhedral signal resampling, the diffusion signal is transformed into a uniform fourth-dimensional tensor of size \(H \times W \times D \times n\), where \(n\) represents the number of resampling directions specific to the chosen icosahedral scheme. For instance, \(n\) is 20 for the icosa6 level, 80 for the icosa21 level, and 320 for the icosa46 level. In the design of the Polyhedra Encoding Transformer, each ``ViT patch" corresponds to a \(3 \times 3 \times 3\) cubic region of diffusion signals along a specific resampling direction. These patches are arranged in a fixed order according to the selected icosahedral scheme, ensuring consistent input structure across varying datasets. 

The patches are then projected into a higher-dimensional embedding space of size 256, followed by the addition of positional embeddings to preserve the spatial context of the input data. The Sinusoidal Positional Encoding~\cite{vaswani2017attention} is then applied to the embeddings. The embedded patches are subsequently processed through a series of six transformer layers. Each transformer layer includes a multi-head self-attention mechanism, which is divided into eight heads, and is followed by a feedforward neural network. To enhance model stability and performance, residual connections and layer normalization are applied to both sub-layers within each transformer layer. The final stage of the model consists of two multi-layer perceptron (MLP) heads, each with a hidden dimensionality of 256. These MLP heads directly produce predictions for the free water fraction and fiber orientation distribution function (FOD) for the central voxel of each input patch. 


\section{EXPERIMENTS}
\label{sec:experiment}

\subsection{Data}
We have chosen diffusion-weighted MRI (DW-MRI) data from the Human Connectome Project - Young Adult (HCP-ya) dataset~\cite{van2013wu, glasser2013minimal}; The Baltimore Longitudinal Study of Aging (BLSA)~\cite{ferrucci2008baltimore}; and the MASiVar dataset~\cite{cai2021masivar}. For HCP-ya, we focusing exclusively on the 1000 \(s/mm^{2}\) b-value shell. A total of 220 subjects were used for the study. The original acquisitions included multiple b-value shells (1000, 2000, 3000 \(s/mm^{2}\)) with 90 gradient directions on each shell, but for this analysis, only the 1000 \(s/mm^{2}\) shell data was utilized. T1-weighted volumes from the same subjects were employed for white matter (WM) segmentation using the SLANT method~\cite{SLANT}. All HCP-ya DW-MRIs underwent distortion correction using top-up and eddy tools~\cite{jenkinson2012fsl}. Out of the 220 subjects, 200 were used for training, while 10 subjects were designated for evaluation and 10 for testing. One hundred and twenty acquisitions from BLSA were used. The BLSA dataset was acquired at a b-value of 700 $s/mm^{2}$ (30 gradient directions) using a Philips 3T scanner. The detailed data collection and preprocessing pipeline can be found at ~\cite{kim2024scalable, kim2024scalable2, cai2021prequal}. Out of the 120 subjects, 100 were used for training, while 10 subjects were designated for evaluation and 10 for testing.

Additionally, the MASiVar dataset was utilized as external validation. Five subjects were acquired on three different sites.  All in-vivo acquisitions were pre-processed with the PreQual pipeline~\cite{cai2021prequal} and then registered pairwise per subject. The acquisitions at b-value of 1000 $s/mm^{2}$ (96 gradient directions) were extracted for the study.

In diffusion MRI, the relationship between the measured signal and the diffusion properties of tissue is described by the Stejskal-Tanner equation. For each gradient direction \(\vec{g}_i\), the signal attenuation is given by:

\begin{equation}
S_i = S_0 \exp\left( -b_i \cdot \text{ADC}_i \right),
\end{equation}

\noindent where \(S_i\) is the signal intensity measured with diffusion weighting along direction \(\vec{g}_i\), \(S_0\) is the signal intensity without diffusion weighting (i.e., the b0 image), \(b_i\) is the diffusion weighting factor, and \(\text{ADC}_i\) is the Apparent Diffusion Coefficient (ADC) along direction \(\vec{g}_i\). By rearranging this equation, we can compute ADC as:

\begin{equation}
\text{ADC}_i = -\frac{1}{b_i} \ln\left( \frac{S_i}{S_0} \right).
\end{equation}

Transforming diffusion MRI data from signal space to diffusion space (i.e., calculating ADC) helps reduce site-specific biases, such as differences in \(B_0\) field strength, which affect signal intensity. This standardization improves the comparability of data across sites. If multiple b0 images are acquired for a voxel, we use the mean signal intensity to compute \(S_0\), ensuring a more robust ADC estimation.

\begin{figure*}[t]
\begin{center}

\includegraphics[width=0.85\linewidth]{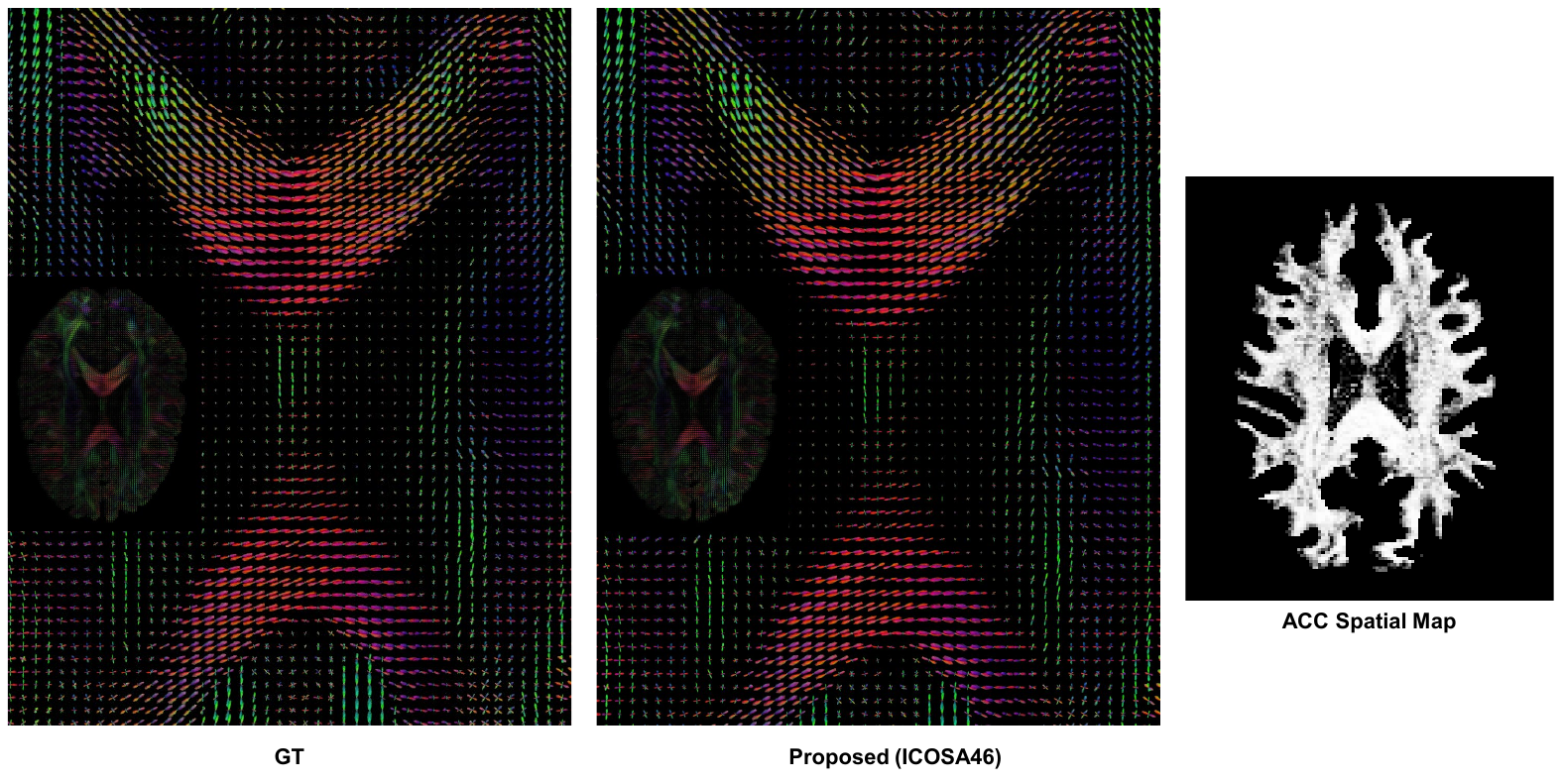}
\end{center}
\caption{\textbf{Qualitative Result.} Two subjects from the test cohort are visualized for assessment of fiber orientation distribution estimation. The ACC spatial map depicts the consistency between the estimation and Ground Truth labels (GT) .}
\label{ODF}
\end{figure*}

\begin{figure*}[t]
\begin{center}
\includegraphics[width=0.85\linewidth]{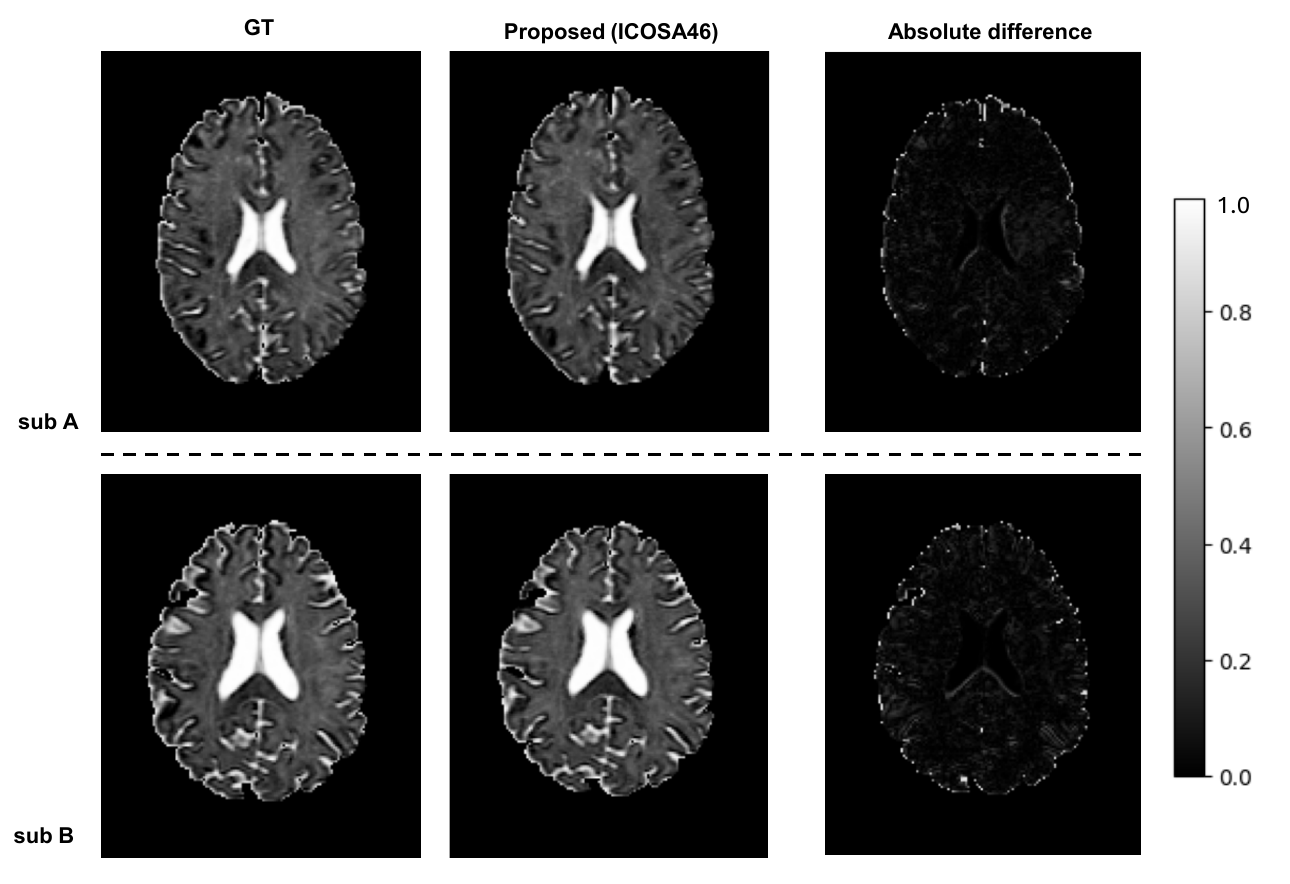}
\end{center}
\caption{\textbf{Qualitative Result.} Two subjects from the test cohort are visualized for assessment of free water estimation.}
\label{FW}
\end{figure*}

\subsection{Baseline methods}
For comparison, we implemented two baseline architectures: a Spherical Harmonics CNN (SH-CNN)~\cite{yao2024robust}, and a vanilla Transformer.

The Spherical Harmonics CNN architecture is designed to take advantage of the compact representation of diffusion signals using spherical harmonics. Specifically, the input to this network is a \(3 \times 3 \times 3 \times 45\) tensor, where 45 represents the number of coefficients derived from the 8th-order spherical harmonics expansion. The network architecture consists of three successive 3D convolutional layers: the first layer (conv1) employs 45 filters with a \(1 \times 1\) kernel size and no padding, followed by a second layer (conv2) with 45 filters and a \(3 \times 3\) kernel size with padding of 1, and a third layer (conv3) with 45 filters and a \(3 \times 3\) kernel size with no padding. To enhance the network's ability to learn deep features, a residual block is included, allowing direct transmission of information from the input to the output of the third convolutional layer via a shortcut connection. This residual connection facilitates the effective training of the deep learning network by mitigating the vanishing gradient problem.

The vanilla Transformer architecture is configured with a maximum sequence length of 200. Since we resampled the signal on both a point and its symmetric counterpart during Polyhedra signal resampling. For the vanilla transformer, the diffusion signals are duplicated and if $\mathbf{\vec{g}}$ is the gradient direction in the original set, then the duplicated signal is assigned to $\mathbf{-\vec{g}}$. This approach allows the model to explicitly learn from both the original and symmetric gradient directions, potentially enhancing its understanding of the diffusion process.  The input consists of \(3 \times 3 \times 3\) patches, which are first flattened and then concatenated with the corresponding \(1 \times 3\) gradient direction, resulting in a \(1 \times 30\) vector. This vector is then linearly projected to form a 256-dimensional embedding. The remaining architecture of the vanilla Transformer mirrors that of the proposed Polyhedra Encoding Transformer, including the addition of positional embeddings, and the processing through a series of six Transformer layers.

\begin{table}[t]
\caption{Free water fraction estimation and fiber ODF estimation on test subjects}
\centering
\begin{tabular}{l|l|l|l|l|l|l|l} 

                             & \multicolumn{2}{l|}{\textbf{HCP}}                         & \multicolumn{2}{l|}{\textbf{BLSA}}                    & \multicolumn{2}{l|}{\textbf{MASIVar}}                     &          \\ 
\hline
                             & FWF                           & SHC                       & FWF                         & SHC                     & FWF                           & SHC                       & p-value  \\ 
\hline
\textbf{Vanila Transformer} & 1.887E-02                   & 0.702                   & 1.857E-02                   & 0.748                   & 1.918E-02                   & 0.731                     & Ref      \\ 
\hline
\textbf{SH-CNN}              & 1.493E-02                   & 0.825                     & 1.546E-02                   & 0.801                   & 1.600E-02                   & 0.805                     &  $p<0.05$  \\ 

\hline
\textbf{Proposed (ICOSA6)}   & 1.819E-02                   & 0.767                     & 1.775E-02                   & 0.776                   & 1.746E-02                   & 0.745                     &  $p<0.05$   \\ 
\hline
\textbf{Proposed (ICOSA21)}  & \textcolor{blue}{ 1.487E-02 } & \textcolor{blue}{ 0.828 } & \textcolor{blue}{1.536E-02} & \textcolor{red}{0.815}  & \textcolor{red}{ 1.594E-02 }  & \textcolor{red}{ 0.816 }  &  $p<0.05$   \\ 
\hline
\textbf{Proposed (ICOSA46)}  & \textcolor{red}{ 1.421E-02 }  & \textcolor{red}{ 0.835 }  & \textcolor{red}{1.507E-02}  & \textcolor{blue}{0.814} & \textcolor{blue}{ 1.597E-02 } & \textcolor{blue}{ 0.810 } & $p<0.05$  \\

\end{tabular}
\label{table:metrics}
\caption*{The top-2 performance is denoted as the \textcolor{red}{red} and \textcolor{blue}{blue} mark, respectively. Statistical assessment is performed via the Wilcoxon signed-rank test.}
\end{table}

\section{RESULTS}
\label{sec:result}  
The results of the Free Water Fraction (FWF) estimation and fiber Orientation Distribution Function (ODF) estimation using spherical harmonics (SHC) across different datasets (HCP, BLSA, and MASIVar) are presented in Table~\ref{table:metrics}. As the ground truth of free water fraction is reconstructed with Free Water Elimination(FWE)~\cite{pasternak2009free} and Constrained Spherical Deconvolution(CSD)~\cite{tournier2007robust}.  The evaluation metrics used are Root Mean Squared Error (RMSE) for FWF, which captures the accuracy of anisotropic feature estimation, and Angular Correlation Coefficient (ACC) for SHC, which assesses the precision of isotropic feature representation.

The proposed Polyhedra Encoding Transformer models, utilizing different levels of icosahedral resampling (ICOSA6, ICOSA21, and ICOSA46), further enhanced performance as compared with the basic transformer, particularly with higher levels of tessellation and resampling. The ICOSA46 variant consistently achieved the best results, with the lowest RMSE values for FWF estimation (e.g., 1.421 E-04 for HCP) and the highest ACC values for SHC (e.g., 0.835 for HCP). The ICOSA21 model also performed robustly, frequently placing in the top two across most metrics and datasets, indicating that higher levels of icosahedral resampling contribute positively to both anisotropic and isotropic feature learning. However, the ICOSA6 variant did not perform as well as the higher-resolution icosahedral configurations. This is likely due to the lower resolution of the q-space sampling inherent in ICOSA6, which consists of only 20 directions. The reduced resolution may not adequately capture the full complexity of the diffusion signals, leading to less accurate feature representation.

\section{CONCLUSION}
\label{sec:conclusion}
In this study, we introduced and evaluated the Polyhedra Encoding Transformer (PE-Transformer), a novel approach for diffusion MRI analysis that leverages icosahedral resampling to achieve uniform and isotropic sampling of q-space data. By comparing this model against baseline architectures, including a Spherical Harmonics CNN and a vanilla Transformer, we demonstrated the effectiveness of our method in accurately estimating both anisotropic and isotropic features of diffusion signals. The Polyhedra Encoding Transformer represents a promising advancement for further development of data-driven deep learning-based dMRI estimator.

\acknowledgments 
This work was supported by the National Institutes of Health under award numbers R01EB017230, 1R01DK135597-01, T32EB001628, K01AG073584 and 5T32GM007347, and in part by the National Center for Research Resources and Grant UL1 RR024975-01. This study was also supported by the National Science Foundation (1452485, 1660816, and 1750213). The Vanderbilt Institute for Clinical and Translational Research (VICTR) is funded by the National Center for Advancing Translational Sciences (NCATS) Clinical Translational Science Award (CTSA) Program, Award Number 5UL1TR002243- 03. The content is solely the responsibility of the authors and does not necessarily represent the official views of the NIH or NSF. This work was also supported by Vanderbilt Seed Success Grant, Vanderbilt Discovery Grant, and VISE Seed Grant. We extend gratitude to NVIDIA for their support by means of the NVIDIA hardware grant. 
\bibliography{report} 
\bibliographystyle{spiebib} 

\end{document}